# Observation of Orbital Hall Effect in Si

Ryoga Matsumoto[1,†], Ryo Ohshima[1,2,†], Mitsuru Funato[1], Yuichiro Ando[1,2,$],

Yuriy Mokrousov[3,4], Dongwook Go[3,4,5,#], and Masashi Shiraishi[1,2,#]

1. Department of Electronic Science and Engineering, Kyoto Univ., 615-8510, Kyoto, Japan.

2. Center for Spintronics Research Network (CSRN), Institute for Chemical Research, Kyoto Univ., 611-0011, Uji, Japan.

3. Peter Grünberg Institut, Forschungzentrum Jülich, 52428 Jülich, Germany.

4. Institute of Physics, Johannes Gutenberg Universität Mainz, 55218 Mainz, Germany.

5. Department of Physics, Korea University, Seoul, 02841, Republic of Korea

[†] These two authors contributed equally to this work.

[#] Corresponding authors: Masashi Shiraishi (shiraishi.masashi.4w@kyoto-u.ac.jp) and

Dongwook Go (dongwookgo@korea.ac.kr)

[$] Present address: Department of Material Science, Osaka Metropolitan Univ., Sakai, Japan.

**Controlling/storing information carriers, such as electron charge and spin, is key for modern information society, and significant efforts have been paid made to establish novel technologies at the nanoscale. The rise of Si-based semiconductor technology and magnetism-based technology has been motivated by the aforementioned demands. However, both technologies have been individually developed, with little effort in fusing them. Hence, establishing a technology to bridge semiconductor and magnetism-based technologies that would allow realization of a novel information device is strongly awaited. In line with this research strategy, the creation of a magnetic device using semiconductors would enable fundamental innovation. Here, we show that a mother material for modern electronics, Si,**

**gives rise to a room-temperature orbital Hall effect (OHE), enabling the creation of novel energy-efficient magnetic memory via efficient torque generation. The orbital torque efficiency $\xi^J_{\mathrm{DL}}$ of Si largely exceeds that of the archetypal metallic materials used in the OHE. Our achievement overturns the conventional understanding that nonmagnetic semiconductors cannot play a pivotal role in magnetic devices and paves a new avenue for creating novel information devices through the fusion of semiconductor and magnetism-based technologies.**

Information-processing devices have been mostly developed with semiconductor transistors using the charge degree of freedom of an electron. However, the spin degree of freedom enables efficient nonvolatile storage of information, which is desirable because information storage via electron charge is energetically inefficient and volatile. Magnetic random-access memory (MRAM), which stores information as the magnetization direction, is an archetypal device used to realize nonvolatile and high-speed memory in information processing using a spin current. Thus, much effort has been made to pioneer an approach that would allow an efficient torque to control spin/magnetization, resulting in a novel magnetization reversal technique with sufficient endurance using the spin Hall effect, the spin-orbit torque (SOT) method. In the SOT method, an electrically generated spin current is utilized for magnetization reversal, through which magnetic bit information can be manipulated. Although spin currents are now playing a pivotal role in modern spintronics/spin-orbitronics and spin current transport has been established in semiconductors, manipulation of spin/magnetization by using nonmagnetic semiconductors is still challenging. In fact, in view of spin current generation, most available semiconductors in the field possess relatively small spin-orbit interactions (SOIs), which hinders efficient creation of a spin current via spin-charge interconversion[1,2] and fusion of semiconductor information-processing and magnetic memory devices.

The orbital angular momentum is one of the most fundamental degrees of freedom in condensed matter physics together with the spin and charge degrees of freedom. In solids, the crystal-field potential tends to strongly suppress the orbital angular momentum, as is known for bulk orbital magnetism. However, the discovery of the orbital Hall effect (OHE)[3–6], which is a superordinate concept of the spin Hall effect (SHE)[5], overturned the conventional understanding because orbital angular momentum flow can be generated by the application of a transverse electric field without the need for SOIs. The SOT due to the SHE is currently the key physical trait for efficient spin-charge interconversion, enabling spin-torque magnetization reversal in MRAMs, where SOIs play a dominant role. In contrast, the OHE does not require SOIs, which is the fundamental difference from the SHE. Although material platforms for efficient SOT have, in principle, been limited to heavy metals with sizable SOIs, the OHE overcomes this limitation and expands the platforms to include light metals, such as Ti[7–9], Cu and Al[10], and the orbital current due to the OHE now enables an efficient torque to manipulate spin information.

Given that the OHE stems from orbital hybridization, the material platforms are not limited to metals but can be expanded to a plethora of materials that possess orbital hybridization. Hence, semiconductors can be novel potential material systems to host the OHE; however, they have not been investigated, i.e., a limited number of theoretical predictions of the OHE focused on p-type Si[3,11] have been reported, and direct experimental demonstration of the OHE in semiconductors is unexplored. Furthermore, although p-type Si can host the OHE stemming from $p$ orbital hybridization in the valence band, substantiating the OHE in p-type Si is not easy since the SHE and OHE in p-type Si are intertwined and are not discernible by conventional experimental procedures, such as THz spectroscopy[12–14] and magnetotransport[8–10,15–20]. In view of the previous

OHE studies and the demand for a novel material platform with high orbital-to-charge conversion efficiency, a material that would allow the creation of magnetic devices with high compatibility with modern electronics and enable the establishment of a sustainable information society is strongly awaited. In this work, we focus on the ubiquitous, nontoxic semiconductors that is most commonly used as a mother material in modern electronics, i.e., silicon (Si), as a platform for observing the OHE by employing phosphorous (P)-doped n-type Si(001) single crystals (Fig. 1). Surprisingly, Si(001) allows the creation of a large OHE, the orbital-to-charge conversion efficiency, $\xi^J_{DL}$, of which largely exceeds that of Ti. Furthermore, first-principles calculation results are consistent with the orbital Hall conductivity obtained from the experiments.

**Orbital-torque generation in Si.**

Figure 2a shows the device structure of a Si-based orbital torque device. the measurement setup, and an optical image of the device. An n-type Si(001) channel (doping concentration of ca. $5 \times 10^{19}$ cm$^{-3}$, i.e. Si was degenerated[21]) with various thicknesses, $t_{Si}$, ranging from 10 to 90 nm on a 200 nm-thick SiO$_2$ insulator was prepared (see also Supplementary Information #1 for the resistivity of the Si channels). Ni was selected as a ferromagnetic metal because of its sizable orbital torque susceptibility[22], and the thickness of Ni, $t_{Ni}$, was varied from 5 to 30 nm. Ni was deposited on Si, and the Ni/Si channel was formed into 15 μm × 100 μm rectangular channels and connected to Au/Ti electrodes by using electron-beam lithography and Ar$^+$-ion milling (see Methods for details). The result of transmission electron microscopy (TEM) is shown in Fig. 2b, where discernible interface formation of the Ni and the Si is confirmed albeit a thin (ca. 1-2 nm) interlayer is partly formed at the interface of the Ni/Si. The possible role of the interlayer to the OHE will be discussed in the next section.

Spin-torque ferromagnetic resonance (ST-FMR) measurements were carried out by injecting an rf electric current with an input power of 100 mW and applying an external magnetic field rotating in the plane of the Ni/Si interface ($\phi$ = 0–360°). The rf current frequency $f$ was changed from 3 to 9 GHz and the external magnetic field was swept from −300 to +300 mT. The electric current injected into the Si layer creates an orbital current via the OHE in the Si, in addition to the Oersted field, and the orbital current is converted into a spin current via SOIs in the Ni layer, resulting in a spin torque. The orbital current from Si due to the OHE and Oersted field enable magnetization precession in Ni, and the oscillation of the resistance of the channel due to anisotropic magnetoresistance results in the DC voltage $V_{DC}$.

Figure 2c shows the DC output voltages under ST-FMR obtained from the Ni(20)/Si(70) device at $\phi$ = 45° as $f$ was swept from 3 to 9 GHz in 0.5 GHz steps (the number in brackets following the stack element is the film thickness in nanometres). A noticeable Lorentzian-shaped DC voltage manifesting successful detection of the orbital torque is observed for the Ni/Si device. Notably, the spin torque in the measurement is generated by the orbital-to-spin conversion in Ni due to orbital flow from Si. The observed DC voltage was fitted by the following equation[23]:

$$V_{DC} = S \frac{\Delta^2}{\mu_0^2(H-H_{res})^2+\Delta^2} + A \frac{\Delta\mu_0(H-H_{res})}{\mu_0^2(H-H_{res})^2+\Delta^2}, \quad (1)$$

where $S$ is the symmetric component, $A$ is the antisymmetric component, $\mu_0$ is the vacuum permeability, $H_{res}$ is the resonance field, and $\Delta$ is the half-width at half-maximum of the Lorentzian function. Whereas the $S$ and $A$ components are attributed to the orbital torque and Oersted field, respectively, other spurious effects, such as the thermal contribution, may superimpose on them. Therefore, the $\phi$ dependences of the $S$ and A components can provide further evidence by decoupling the spurious contributions from the orbital torque, where the $\phi$ dependences of $S$ and $A$ are described as [24,25]:

$$S = S_0 \sin 2\phi \cos \phi + S_1 \sin 2\phi + S_2 \sin 2\phi \sin \phi + S_3 \sin \phi, \quad (2a)$$

$$A = A_0 \sin 2\phi \cos \phi + A_1 \sin 2\phi + A_2 \sin 2\phi \sin \phi, \quad (2b)$$

where $S_0$ is the component of the anisotropic magnetoresistance due to spin torque from the spin angular momentum $\sigma_y$ i.e. damping-like (DL) torque attributed from the OHE, and $A_0$ is the component of field-like torque due to $\sigma_y$ and the $y$ component of the Oersted field. Other parameters, $S_1$, $S_2$, $S_3$, $A_1$, and $A_2$ are the components of the anisotropic magnetoresistance due to different damping-like and field-like spin torques from the spin angular momentum $\sigma_{x/z}$, the spin-pumping-induced inverse SHE and anomalous Nernst effect due to the thermal gradient in the device[24,25]. Figures 2d and 2e show the $\phi$ dependences of the $S$ and $A$ components obtained for the Ni(20)/Si(70) device, respectively. The $S_0$ component is dominant in the Ni(20)/Si(70) device, which unequivocally affirms that the observed spin torque is ascribed to the OHE in the n-type Si, as the SHE is negligible because of the small SOIs in n-type Si[26,27] (see also Supplementary Information #3 for the $\phi$ dependences of $S$ and $A$ for Si and Ni with various thicknesses).

The DL torque efficiency per electric field, $\xi^E_{DL}$ as an index of the strength of the OHE, was estimated from the obtained $S_0$ as follows[8,19]:

$$\xi^E_{DL} = \frac{4eM_S t_{FM} S_0 \Delta}{\hbar E I_{rf} R_{AMR}} \frac{2H_{res}+M_{eff}}{H_{res}+M_{eff}} \sqrt{1+\frac{M_{eff}}{H_{res}}}, \quad (3)$$

where $e$ is the elemental charge, $\hbar$ is the Dirac constant, $M_{eff}$ is the effective saturation magnetization estimated from the frequency dependence of $H_{res}$ (see Supplementary Information #2), $E$ is the applied electric field, $I_{rf}$ is the amplitude of the injected rf current, and $R_{AMR}$ is the amplitude of the anisotropic magnetoresistance (see Supplementary Information #4 for details). The $\xi^E_{DL}$ of the Ni(20)/Si(70) channel was estimated to be 270 $(\Omega \cdot cm)^{-1}$ from the calculation using Eq. (3) with averaging the amplitudes of three different samples. The comparable torque efficiency in Ni(20)/Si(70) was corroborated as well via the second harmonic Hall measurement (see Supplementary Information #5 for details). To compare the torque efficiency with that of materials previously investigated, the torque efficiency in n-Si normalized by the injected current density, i.e., $\xi^J_{DL}$ (the effective orbital Hall angle $\theta_{eff}$) = $\xi^E_{DL}\rho_{NM}$, where $\rho_{NM}$ is the longitudinal electrical resistivity of the nonmagnetic layer[8,20,28], was estimated. Since 20 nm-thick single Ni layer shows the self-induced spin-orbit torque (see the discussion in the next section and Supplementary Information #6), the effective orbital torque efficiency $\xi^J_{DL,eff}(t_{Si})$ = ($\xi^E_{DL}(t_{Si})$ − $\xi^E_{DL}(0))\rho_{NM}$ is introduced to discuss the contribution of the orbital Hall effect in Si. Surprisingly, the $\xi^J_{DL,eff}$ of n-Si is extremely large and has a maximum value of 1.5, largely exceeding that of Ti (~ 0.1) (see Supplementary Information #7 for the estimation of $\xi^J_{DL,eff}$ for Si channels with various thicknesses). Furthermore, given that Ti has one of the highest torque efficiencies among previous OHE studies, this comparison strongly highlights the great potential of Si as an orbital-to-spin conversion material.

**Thickness dependence of the orbital torque efficiency.**

To delve further insight into the microscopic mechanism of the OHE in n-Si, the Si thickness ($t_{Si}$) dependence of the $\xi^E_{DL}$ of Ni(20)/Si($t_{Si}$) was investigated. $\xi^E_{DL}$ monotonically increases with increasing the Si up to 30 nm and then saturates (see Figs. 3a and 3b). The $\xi^E_{DL}$ with Si layers (especially >30 nm) discernibly surpasses the $\xi^E_{DL}$ of the single Ni layer (Ni(20)/Si(0)), highlighting the contribution to the spin-orbit torque from Si (as an additional experiments, the second harmonic Hall measurement of the single Ni layer grown on SiO$_2$(500 nm) was implemented, where indiscernible magnetic field angular dependence compared with Ni(20)/Si(70) devices was observed, resulting in difficulty of the estimation of damping-like torque in a single Ni layer, see Supplementary Information #6). To note as well, absence of the OHE in Ni/SiO$_2$/Si multilayer devices signifies that the Si allows the orbital torque generation (see Supplementary Information #8). The change in the effective magnetization of Ni, $M_{eff}$, is indiscernible in the examined range of $t_{Si}$, indicating that the magnetization of Ni does not change when the Si film thickness is changed by applying a milling process on Si (see Fig. 3c and Methods). Hence, the results shown in Figs. 3b and 3c indicate that the change in the orbital toque of Si as a function of its thickness induces a change in $\xi^E_{DL}$ and simultaneously negates interfacial effects such as orbital/spin Rashba effects[29–31] and alloying[32,33], i.e., the interlayer observed in the TEM viewgraph does not play a crucial role to the OHE in Si. The experimental data were fitted by the conventional function[8], $\xi^E_{DL}(t_{Si}) = \xi^E_{DL,0}(1 - \text{sech}(t_{Si}/\lambda_{Si})) + \xi^E_{DL}(0)$, where the first and second terms describe the orbital torque contribution due to the OHE in n-Si and the self-induced spin-orbit torque contribution in single Ni layer, respectively. With this consideration, we obtain $\xi^E_{DL,0} = 130 \pm 30$ $(\Omega \cdot cm)^{-1}$ and $\lambda_{Si} = 12 \pm 8$ nm as the thickness-independent torque efficiency and the orbital decay length of n-Si, while $\xi^E_{DL}(0) = 210$ $\Omega^{-1}cm^{-1}$ is taken from the measurement on the single Ni film. We note that the orbital decay lengths of nonmagnets in previous studies using metals ranged from a few nanometres[9,17,19] to 70 nm[7,8], and the orbital decay length of Si is comparable to that of Ti.

Other significant supporting evidence was obtained from the Ni thickness ($t_{Ni}$) dependence of the $\xi^E_{DL}$ of Ni($t_{Ni}$)/Si(70). $\xi^E_{DL}$ monotonically increases with increasing Ni thickness up to 30 nm (see Figs 4a and 4b), which is consistent with the long-range orbital current propagation in Ni experimentally and theoretically reported in previous studies[8,34]. Notably, the weak decrease in $\mu_0 M_{eff}$ in the thinner Ni may be ascribed to the surface magnetic anisotropy. A monotonic increase in the $\xi^E_{DL}$ of Ni was reported in a previous study[8]. This fact is a distinct manifestation of orbital current transport in Ni from Si because $\xi^E_{DL}$ does not exhibit a thickness dependence unless orbital current transport occurs, i.e., a spin current transport is transported from Si due to the SHE given that the spin diffusion length of Ni at room temperature is 3.3 nm as reported in Ref. 35.

**First-principles calculation of the OHE in electron-doped Si.**

The experimental observation of a sizable DL torque efficiency, $\xi^E_{DL} = +270\ (\Omega\cdot cm)^{-1}$, is a positive sign of angular momentum flux into Ni. However, a previous computational study by Yao and Fang[36] predicted a negative sign for the SHE in electron-doped Si, which cannot account for our experimental results. This discrepancy suggests the presence of another source of nonequilibrium angular momentum, which likely originates from the orbital, rather than spin, degree of freedom. The OHE in hole-doped Si was theoretically investigated by Bernevig *et al.*[3], who identified *p-p* orbital hybridization near the valence band top at Γ as the microscopic origin, predicting a positive sign. However, a detailed theoretical investigation of the OHE in electron-doped Si is lacking.

Motivated by this, we conducted an in-depth analysis of the origin of the OHE in electron-doped Si using a recently proposed gauge-covariant formalism[37] (see Methods for computational details). First, we examined the orbital character of the band structure to determine whether the conduction band has sufficient *p* orbital character to support angular momentum transport under nonequilibrium conditions. As shown in Fig. 5a, the conduction band minimum of Si is located at the X point, where the orbital character exhibits an admixture of *s* and *p* orbitals. This behaviour contrasts with that of GaAs, where the conduction band minimum is at the Γ point and exhibits a purely *s* orbital character[38]. Consequently, in Si, the momentum-space orbital texture of conduction electrons is shaped by *s-p* and *p-p* orbital hybridizations, which facilitate the generation of nonequilibrium orbital angular momentum under an external electric field[6].

Figure 5b shows the computed orbital Hall conductivity ($\sigma_{OHE}$) and spin Hall conductivity ($\sigma_{SHE}$) as functions of the carrier concentration ($n_e$). The main prediction is that $\sigma_{OHE}$ has a positive sign, which is consistent with the experimental results. In addition to the results of the OHE in n-type Si, the orbital torque detection in p-type Si with the doping concentration of $3\times10^{19}$ cm$^{-3}$ is also confirmed with ST-FMR measurement (see Supplementary Information #9), where the efficiency is estimated to be 120 $\Omega^{-1}$cm$^{-1}$. As shown in Fig. 5b, the amplitude of the orbital Hall conductivity of degenerated p-Si is comparable to that of n-Si, which is consistent with the experimental findings. Here, the calculation result differs from the negative sign predicted by Baek and Lee[11]. We clarified that this discrepancy arises from the anomalous position correction, which stems from orbital dipole matrix elements[37] (see also Supplementary Information #10 for the details). In contrast, $\sigma_{\text{SHE}}$ is significantly smaller than $\sigma_{\text{OHE}}$ because of the negligible SOIs in Si. Notably, $\sigma_{\text{SHE}}$ changes sign from positive to negative when transitioning from the hole-doped to electron-doped regime. Interestingly, we also find that Si behaves as an orbital Hall insulator, a feature recently identified in several two-dimensional materials[39–41]. Although the role of the so-called undergap orbital current[42] in current-induced torques remains unclear, we speculate that this current may be absorbed by ferromagnetic states if they align with the undergap states in Si. Regardless of the undergap orbital current, even when the offset value of the gap is set to zero, our calculations still predict a positive sign for the OHE and a negative sign for the SHE in electron-doped Si.

**Summary and outlook**

In summary, we successfully observed a orbital Hall torque from n-Si that does not inherently possess SOIs giving rise to a spin torque. Our assertion was supported by systematic experiments, i.e., ST-FMR and second harmonic Hall measurements, and by the Si and Ni thickness dependences of the orbital torque. The magnitude of the orbital torque efficiency $\xi^J_{\text{DL}}$ largely exceeds those in conventionally investigated material systems, which allows us to envisage various applications. Given the high orbital torque efficiency for the low longitudinal conductivity in Si, smaller electric current injection would allow an efficient orbital/spin torque, giving rise to energy-efficient magnetic memory devices. Furthermore, the combination of a Si OHE channel and a diluted ferromagnetic semiconductor[42,43] enables all-semiconductor magnetic memory devices with sufficient energy efficiency and material ubiquity that can operate at room temperature. In addition, given that crystallization is not always necessary in the creation of a sizable orbital current[44], the introduction of amorphous Si as the OHE channel may be realized, which would allow flexible magnetic memory devices. Thus, our achievement provides fascinating prospects for novel information technologies such as the fusion of semiconducting and magnetic traits in condensed matter.

**Method**

**Device fabrication**

A commercially available non-doped 100 nm-thick Si(001) on $SiO_2$(200 nm)/Si wafer (SOITEC) was cut into 1 inch-square pieces for the ion implantation and cut into 5 mm-square for the device fabrication. A 100 nm-thick n-type Si channel (doping concentration was ca. $5 \times 10^{19}$ cm$^{-3}$, that is, Si was degenerated[21]) on a 200-nm-thick $SiO_2$ insulator was prepared by implanting phosphorous into Si by using an ion implanter at various acceleration voltages (NISSIN ELECTRIC) and annealing at 1000°C for 30 s by using a rapid-thermal annealing system (Heatpulse 610, AG Associates). An almost uniform resistivity profile at various Si depths was confirmed by $Ar^+$-ion milling (Hakuto) and resistivity measurement of samples with various Si thicknesses via the van der Pauw method (see Supplementary Information #1). The Si surface was milled by $Ar^+$-ion milling to obtain the target Si thickness, and Ni was deposited by using electron-beam (EB) deposition. The Ni/Si channel was formed into 15 μm × 100 μm rectangular channels and connected to Au(100)/Ti(3) electrodes as a part of the signal line of a coplanar waveguide consisting of the Ni/Si channel by using EB lithography and $Ar^+$-ion milling.

**ST-FMR measurement**

The ST-FMR measurement was carried out by injecting an rf current generated by a signal generator (Keysight) with an input power of 100 mW and applying an external magnetic field rotating in the plane of the Ni/Si interface ($\phi$ = 0–360°). The rf current frequency $f$ ranged from 3 to 9 GHz and the external magnetic field was swept from −300 to 300 mT. The Oersted field and the orbital current from Si due to the OHE induced magnetization precession of Ni, and the oscillation of the resistance of the channel due to the anisotropic magnetoresistance of Ni resulted in the DC voltage $V_{DC}$ detected by a nanovoltmeter (Keithley). All the measurements were carried out at room temperature.

**First-principles calculation**

For the calculation of the orbital and spin Hall conductivities in Si, we employed the Wannier interpolation technique. We performed self-consistent density functional theory calculations via the full-potential linearly augmented plane wave method[46], implemented in the FLEUR code[47]. We used the Perdew-Burke-Ernzerhof functional in the scheme of the generalized gradient approximation for the exchange and correlation effects[48]. The lattice constant of Si was set to $a$ = 10.26$a_0$, where $a_0$ is the Bohr radius. The muffin-tin radius was set to 2.16$a_0$, in which the harmonic expansion up to $l_{max}$ = 12 was included. The plane wave was expanded in the interstitial region up to $K_{max}$ = 5.0/$a_0$. Based on the converged Kohn-Sham states, we constructed maximally localized Wannier functions by using the WANNIER90 code[49]. We used a total of 16 $sp_3$ orbitals for spin-up and spin-down states centred on two Si atoms as initial projections and minimized the spread, for which the maximum of the frozen energy window in the disentanglement was set to 4 eV above the conduction band bottom. Based on the Wannier representations of the Hamiltonian, position, and orbital angular momentum operators, the orbital Hall conductivity was computed by evaluating the Kubo formula,

$$\sigma_{\text{OHE}} = -\frac{e\hbar}{2} \int \frac{d^3k}{(2\pi)^3} \sum_{nn'} (f_{n\bm{k}} - f_{n'\bm{k}}) \frac{\text{Im}[\langle\psi_{n\bm{k}}|(v_z L_y + L_y v_z)|\psi_{n'\bm{k}}\rangle\langle\psi_{n'\bm{k}}|v_x|\psi_{n\bm{k}}\rangle]}{(E_{n\bm{k}} - E_{n'\bm{k}})^2 + \eta^2},$$

in a homebuilt code, where $e > 0$ is the unit charge, $\hbar$ is the reduced Planck constant, $\psi_{n\bm{k}}$ is the Bloch state with band index $n$ and crystal momentum $\bm{k}$, $E_{n\bm{k}}$ is its energy eigenvalue, $f_{n\bm{k}}$ is the corresponding Fermi-Dirac distribution function, and $L_y$ is the $y$ component of the orbital angular momentum operator, which is integrated in the muffin-tin sphere (atom-centred approximation). Importantly, $v_\alpha$ is the $\alpha$ component of the velocity operator with anomalous position correction, given by

$$v_\alpha = \frac{1}{\hbar}\frac{\partial H(\bm{k})}{\partial k_\alpha} + \frac{1}{i\hbar}[A_\alpha(\bm{k}), H(\bm{k})],$$

where $H(\bm{k})$ is the Hamiltonian, and $A_\alpha(\bm{k})$ is the $\alpha$ component of the anomalous position. The anomalous position can be obtained from the dipole matrix elements between Wannier functions[37]. We set the temperature $T = 300$ K in the Fermi-Dirac distribution function, and used the broadening parameter $\eta = 25$ meV. The $\bm{k}$-points were sampled over a 512×512×512 mesh grid in the first Brillouin zone.

**Acknowledgements**

R.M., R.O, Y.A, and M.S. acknowledge experimental support by Motomi Aoki and Katsuhiro Tatsuoka. D.G. acknowledges fruitful discussion with Insu Baek and Hyun-Woo Lee. A part of this work was supported by the "Advanced Research Infrastructure for Materials and Nanotechnology in Japan (ARIM)" of the Ministry of Education, Culture, Sports, Science and Technology (MEXT) and the MEXT Initiative to Establish Next-Generation Novel Integrated Circuits Centers (X-nics, Tohoku University, Japan), Grant Number JPMXP1223NU0230. D.G.

and Y.M. are supported by the Deutsche Forschungsgemeinschaft (DFG, German Research Foundation)−TRR 173/3−268565370 (project A11). D.G. and Y.M. also acknowledge support from EIC Pathfinder OPEN grant 101129641 "OBELIX". D.G. and Y.M. also gratefully acknowledge the Jülich Supercomputing Centre for providing computational resources under project jiff40.

**Author contribution**

M.S. and R.O. conceived the experiments. R.M. and M.F. fabricated samples and collected data. R.O., R.M. Y.A. and M.S. analyzed the experimental results. D.G. and Y.M. implemented theoretical calculations. M.S., R.O. and D.G. wrote the paper. All authors discussed the results.

**Data availability**

The data that support the findings of this study are available from the corresponding author upon reasonable request.

**Supplementary Information**

No. 1 Thickness dependence of the Si resistivity

No. 2 Frequency dependence of the ST-FMR spectra

No. 3 ST-FMR spectra and their angular dependence

No. 4 Estimation of the DL torque efficiency per electric field

No. 5 Second harmonic Hall measurement of Ni/Si device

No. 6 Investigation of the spin-orbit torque in a single Ni layer device

No. 7 Estimation of $\zeta^J_{DL}$

No. 8 ST-FMR and second harmonic Hall measurements with a Ni/SiO$_2$/Si device

No. 9 ST-FMR measurement with a Ni/p-Si device

No. 10 Comparison of theoretical calculation of orbital Hall conductivity with previous studies ferromagnetism in n-type ferromagnetic semiconductor (In,Fe)As grown on vicinal GaAs substrates. *Jpn. J. Appl. Phys.* **59**, 063002 (2020).

44. Goel, S. *et al.* Room-temperature spin injection from a ferromagnetic semiconductor. *Sci. Rep.* **13**, 2181 (2023).

45. Gao, T. *et al.* Control of dynamic orbital response in ferromagnets via crystal symmetry. *Nat. Phys.* **20**, 1896–1903 (2024).

46. Wimmer, E., Krakauer, H., Weinert, M. & Freeman, A. J. Full-potential self-consistent linearized-augmented-plane-wave method for calculating the electronic structure of molecules and surfaces: $O_2$ molecule. *Phys. Rev. B* **24**, 864–875 (1981).

47. Wortmann, D. *et al.* FLEUR. zenodo.

48. Perdew, J. P., Burke, K. & Ernzerhof, M. Generalized Gradient Approximation Made Simple. *Phys. Rev. Lett.* **77**, 3865–3868 (1996).

49. Mostofi, A. A. *et al.* An updated version of wannier90: A tool for obtaining maximally-localised Wannier functions. *Comput. Phys. Commun.* **185**, 2309–2310 (2014).

**Figures & captions**

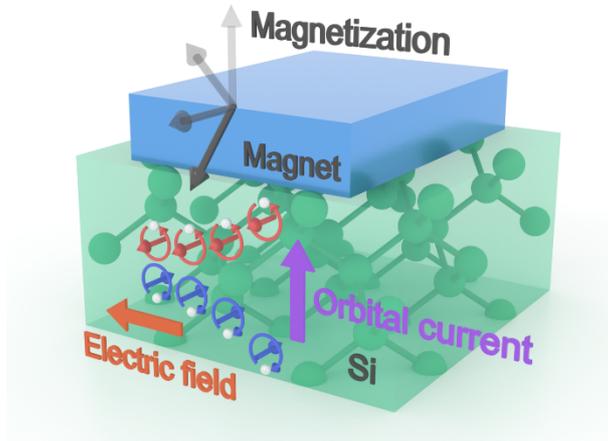

**Fig. 1: Schematic illustration of the OHE in silicon.** An external electric field in silicon generates a transverse orbital current via the OHE, which exerts a torque on the magnetization of an adjacent magnet.

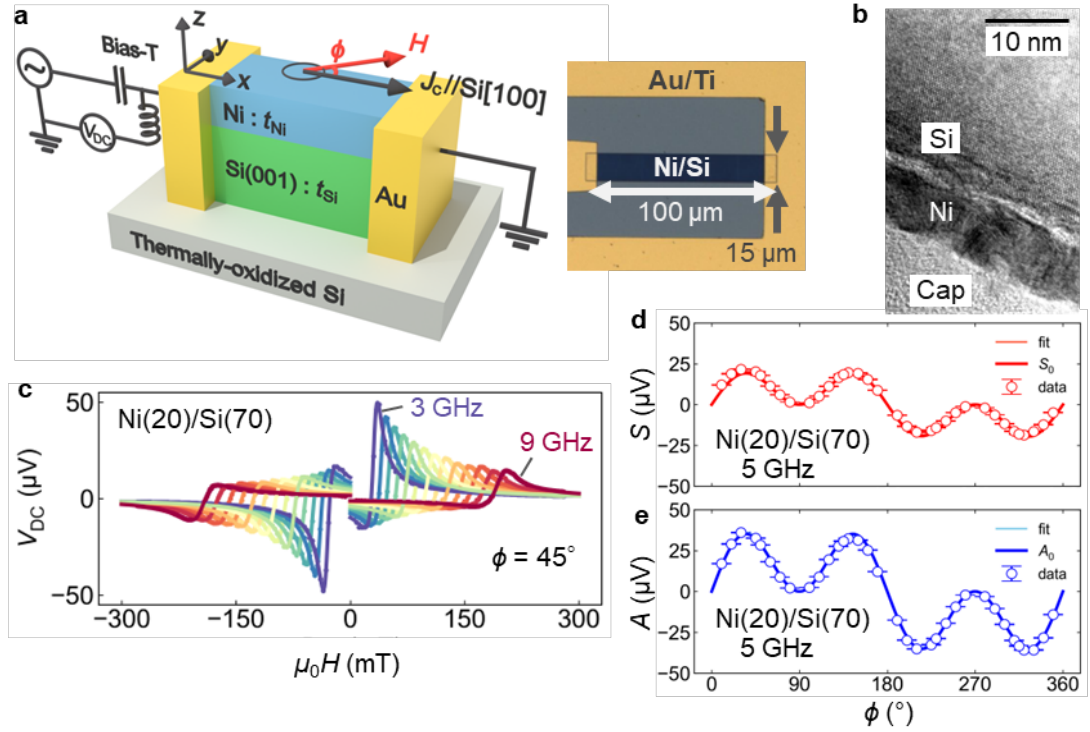

**Fig. 2: ST-FMR in Ni/Si. a.** Schematic illustration of the ST-FMR measurement setup for the Ni/Si(001) device and optical image of the device. The channel direction is set to Si[100]. An rf electric current is injected into Ni/Si, and a DC output voltage is measured. The external magnetic field is rotated in the plane of the Ni/Si film. **b**. A transmission electron microscopy (TEM) image of Ni/Si interface. The capping layer is MgO. **c.** DC voltages obtained from Ni(20)/Si(70) with a 100 mW microwave power input when the angle of the external magnetic field was set to 45°. The microwave frequency $f$ ranged from 3 to 9 GHz. The solid lines are the fitting results obtained via Eq. (1). Magnetic field angular dependence of the **d.** symmetric component $S$ and **e.** antisymmetric component $A$ at $f$ = 5 GHz. The open circles and solid lines are the data points and the fitting results obtained by using Eq. (2), respectively. The bold solid lines indicate the $S_0$ and $A_0$ components in Eq. (2), respectively. The error bar for each data point indicates the standard error.

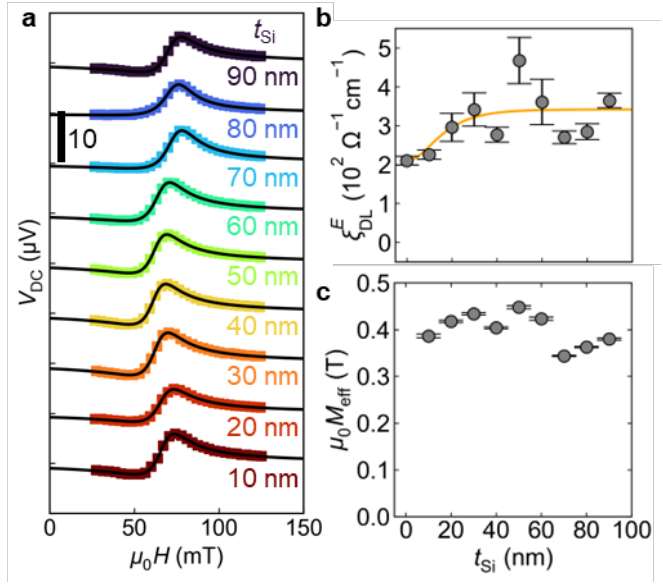

**Fig. 3: Si thickness dependence of the orbital torque efficiency per electric field. a.** DC voltages in Ni(20)/Si($t_{Si}$) with different Si thicknesses ranging from 10 nm to 90 nm at $\phi = 45°$ and $f = 5$ GHz. Salient output voltages are observed. The solid lines are lines fitted via Eq. (1). Si thickness dependence of **b.** $\xi^E_{DL}$ and **c.** $\mu_0 M_{eff}$. The solid line in **b.** indicates the fitting result obtained by using equation of $\xi^E_{DL}(t_{Si})$ in the main text. Averaged data for several devices in the same design are plotted, where the error bars representing the standard error are also averaged by the room mean square. As shown in **c.** $\mu_0 M_{eff}$ is almost unchanged with increasing Si thickness.

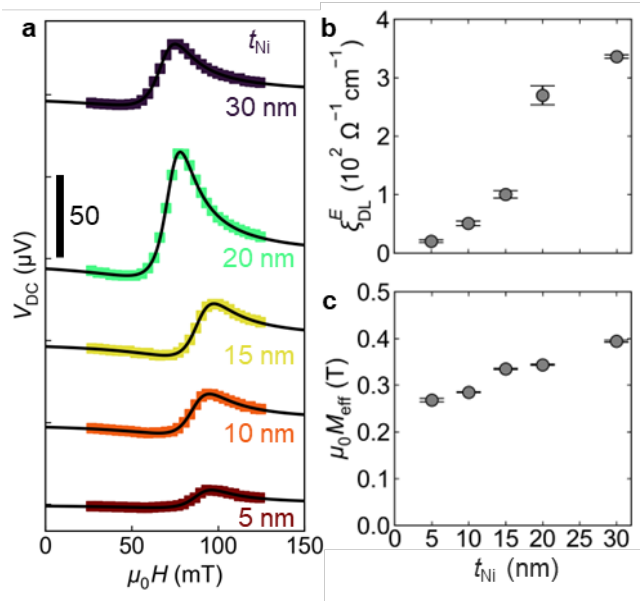

**Fig. 4: Ni thickness dependence of the orbital torque efficiency per electric field. a.** DC voltages from Ni($t_{Ni}$)/Si(70) with different Ni thicknesses ranging from 5 nm to 30 nm at $\phi = 45°$ and $f = 5$ GHz. The solid lines are fitting lines obtained via Eq. (1). Ni thickness dependences of **b.** $\xi^E_{DL}$ and **c.** $\mu_0 M_{eff}$. Averaged data for several devices in the same design are plotted, where the error bars representing the standard error are also averaged by the room mean square. The gradual decrease in the $\mu_0 M_{eff}$ of Ni with decreasing $t_{Ni}$ is discussed in the main text.

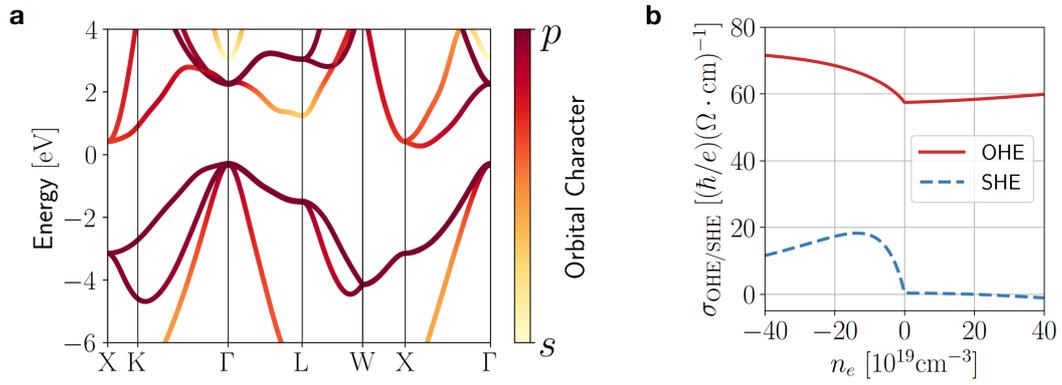

**Fig. 5: First-principles calculations for Si. a**. Band structure along high-symmetry lines with the orbital character weight indicated by colour. **b**. Orbital Hall conductivity $\sigma_{\text{OHE}}$ (red solid line) and spin Hall conductivity $\sigma_{\text{SHE}}$ (blue dashed line) as functions of carrier concentration n$_e$. Positive and negative values of $n_e$ correspond to the electron and hole doping regimes, respectively.